\documentclass[aps, prl, reprint, nofootinbib, superscriptaddress, noshowpacs,floatfix]{revtex4-2}
\usepackage{amssymb}
\usepackage{amsmath}
\usepackage{graphicx}
\usepackage{bm}
\usepackage{lmodern}
\allowdisplaybreaks

\graphicspath{ {./figures/} }

\renewcommand\vec{\mathbf}

\newcommand{\ud}{\,{\mathrm{d}}}
\newcommand{\ukB}{k_\mathrm{B}}

\begin{document}
\title{Switching of a closed mobile vacancy based memristor, whose specific resistance linearly depends on local vacancy concentration}
\author{\firstname{I.~V.} \surname{Boylo}}
\email[]{boylo@donfti.ru}
\affiliation{Galkin Donetsk Institute for Physics and Engineering, R.~Luxembourg str.~72, Donetsk 283048, Russian Federation}
\author{\firstname{K.~L.} \surname{Metlov}}
\email[]{metlov@donfti.ru}
\affiliation{Galkin Donetsk Institute for Physics and Engineering, R.~Luxembourg str.~72, Donetsk 283048, Russian Federation}
\date{\today}
\begin{abstract}
The linear (proportional to local vacancy concentration) term in specific resistance of the material does not directly contribute to the change of memristor's total resistance when the vacancies are redistributed inside while keeping their total number constant. But it still changes kinetics of the vacancy drift under the influence of a passing electric current. These changes are especially significant in the presence of metal-insulator phase transition in the memristor's material. In this paper, kinetic equation for local vacancy concentration is obtained, and exact solutions for its steady states are analyzed. It is shown that not only in the weakly nonlinear case (when the dependence of the specific resistance on the vacancy concentration can be neglected), but also in a strongly non-linear memristor with phase transition, its kinetics can be reduced to the classical exactly solvable Burgers equation.
\end{abstract}
\maketitle
\section{Introduction}
Memristor is a two-terminal element of electric circuits proposed by Leon Chua~\cite{Chua1971,Chua1976}. Its resistance depends on the total electric charge passed through. Such an element was implemented by Hewlett Packard based on the migration of charged oxygen vacancies in titanium dioxide~\cite{strukov2008}. It can also be made based on other materials that allow the migration of ions~\cite{Waser2007} --- various transition metal oxides~\cite{Sawa2008}, manganites~\cite{Bryant2011,Yao2017}. The main property of a memristor is the presence of hysteresis on its volt-ampere characteristic. Hysteresis occurs because the state of the memristor is determined not only by the instantaneous value of the passing electric current, but also by its past values (i.e. by the history of external influences). In a mobile vacancies based memristor, the state is represented by a certain spatial distribution of the vacancy concentration, and its change occurs through diffusion, controlled by an external electric current. The measured resistance of the memristor to electric current reflects the instantaneous distribution of vacancies.

In the literature, the state change of such memristors is modeled (mainly numerically) by various kinetic equations for the vacancy concentration~\cite{agudov2020,agudov2021,roldan2023}. If the specific resistance of the material is constant, nonlinear kinetics of the memristor can be mapped to the Burgers equation. This allows to obtain exact analytical solutions for both transient~\cite{boylo2020} and periodic~\cite{BM2024-en} processes taking place when a current meander is passed through a closed memristor. The change in the resistance of such a memristor (with constant specific resistance of the material) is entirely determined by interfacial effects and depends on the concentration of mobile vacancies in the vicinity of its contacts~\cite{boylo2020}.

If the specific resistance of the memristor material $\rho$ depends on the concentration of vacancies $c$ linearly: $\rho=a + \beta\cdot c$, where $a, \beta=const$, the variable component of the total resistance of the closed memristor is still determined exclusively by interfaces. The volume contribution to the resistance depends only on the total number of vacancies, and therefore in a closed memristor it does not depend on their (determined by the history of external influences) spatial distribution. But kinetics of the vacancies redistribution still depends on the coefficient $\beta$ and under certain conditions this dependence can manifest itself quite significantly. It is these effects, interesting from our point of view, that are studied in detail in the present paper.

 \section{Model}
 Consider a thin film of thickness $d$ made of a material with charged mobile vacancies, bounded by two impermeable to vacancies flat metal contacts. Let's chose the coordinate $x$ to specify the position along the film thickness $0\le x \le d$. The state of such a memristor at time $t$ is described by the instantaneous local concentration of mobile vacancies $C(x,t)$. Moreover, their total number $\int_0^d C \ud x$ is constant, independent of time. Due to the local conservation of the number of vacancies, their motion obeys the continuity equation
 \begin{equation}
  \label{eq:continuity}
 \partial_t C(x,t) + \vec{\nabla}\cdot\vec{J}(x,t) = 0,
 \end{equation}
where $\partial_t$ is the time derivative. In the one-dimensional case, the vacancy flux $\vec{J}=\{J,0,0\}$, and $\vec{\nabla}\cdot\vec{J}(x,t)=\partial_x J(x,t)$. Considering the diffusion and thermally activated hopping transport of vacancies under the action of electric current, we can represent~\cite{boylo2020}
\begin{equation}
 \label{eq:current}
 J = -D\partial_x C + C (1-C/C_\mathrm{max}) \frac{2 D}{a}\sinh\frac{a q \rho I}{\ukB T},
\end{equation}
where $D$ is the diffusion coefficient, $q$ is the charge of one vacancy, $I$ is the current passed through the memristor, $\ukB$ is the Boltzmann constant, $T$ is the temperature, $a$ is the vacancy jump length (the distance between adjacent equilibrium vacancy positions in the material), $\rho$ is the specific resistance of the material. Here, it is also assumed that the local concentration of vacancies is limited by some predetermined value $C_\mathrm{max}$, which is a property of the material~\cite{boylo2020}.

Let us pass to the dimensionless coordinate $\xi=x/d$, dimensionless time $\tau=t D/d^2$ and introduce the normalized concentration of vacancies $c=C/C_\mathrm{max}$, $0\le c(\xi,\tau) \le 1$. Then the linear dependence of the specific resistance of the material on the vacancy concentration can be conveniently represented as
\begin{equation}
 \label{eq:rho}
 \rho = \rho_0 (1 + \beta (c(\xi,\tau) - r)),
\end{equation}
where the memristor fill factor $r$ is
\begin{equation}
\label{eq:r}
 0 < r=\int_0^1 c(\xi,\tau) \ud \xi = const < 1,
\end{equation}
and $\rho_0 = \int_0^1 \rho \ud \xi = const$ --- average specific resistance of the memristor material. The parameter $\beta$ characterizes the strength of the influence of the vacancy concentration on the specific resistance and can be either positive or negative. Substituting~\eqref{eq:rho} into~\eqref{eq:current} and~\eqref{eq:continuity}, passing to the limit of continuous drift $a\rightarrow 0$ for the normalized vacancy concentration $c$, we obtain the equation
\begin{equation}
 \label{eq:kinetics}
 \partial_\tau c + p \left(1\!-\!r\beta\!-\!c \left(2 \!-\!2 (1\!+\!r)\beta\!+\!3\beta c\right)\right)\partial_\xi c - \partial_{\xi\xi}c = 0,
\end{equation}
where the parameter $p=2 q d \rho_0 I/(\ukB T)$ plays the role of the dimensionless effective value of the current passing through the memristor. In the limit $\beta\rightarrow 0$, this equation reduces to the classical Burgers equation~\cite{boylo2020}, but in the general case contains a higher (3rd) order nonlinearity in $c$. The problem of the kinetics of the considered closed memristor reduces to solving this equation with the boundary conditions $J|_{x=0} = J|_{x=d} =0$.

Interestingly, the solutions of the equation~\eqref{eq:kinetics} for positive and negative values of the parameter $\beta$ are related by the transformation
\begin{equation}
 \label{eq:betasign}
 \beta\rightarrow-\beta,\quad
 p\rightarrow-p,\quad r\rightarrow1-r,\quad c\rightarrow1-c.
\end{equation}
Thus, it is enough to solve~\eqref{eq:kinetics} only for positive (or only for negative) values of $\beta$. The solutions with $\beta$ of the opposite sign can then be obtained by simply redefining the parameters according to~\eqref{eq:betasign}. All consequences from the equation~\eqref{eq:kinetics} are transformed in the same way.

\begin{figure}[htp]
\begin{center}
\includegraphics[width=1.0\columnwidth]{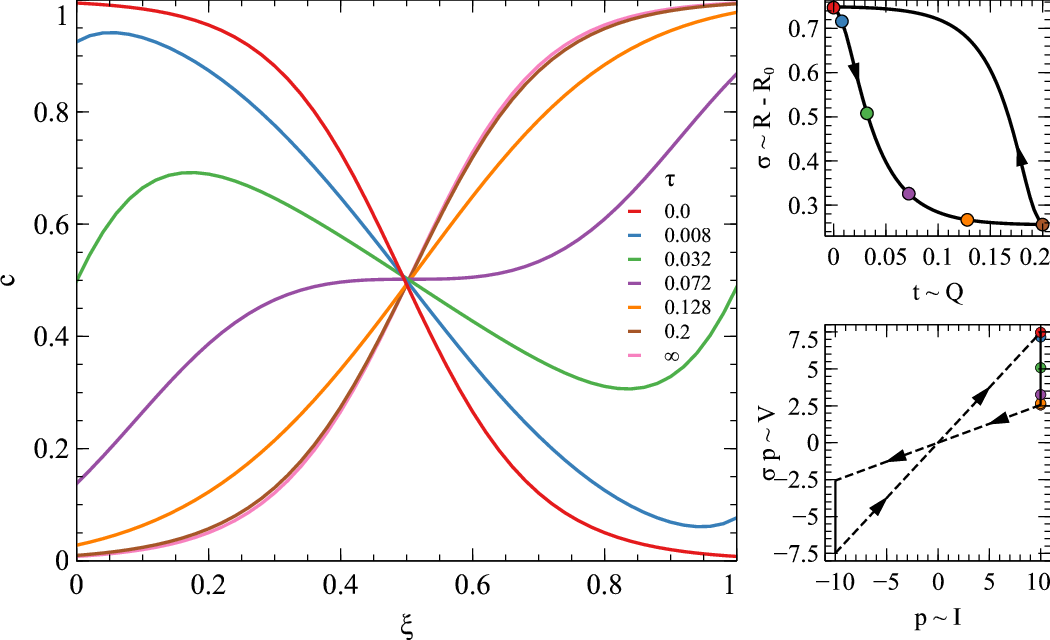}
\vspace{-0.8cm}
\end{center}
\caption{\label{fig:profiles} (left) The concentration profiles $c(\xi,\tau)$ obtained by numerical solution of the equation~\eqref{eq:kinetics} at different times $\tau$ when switching a memristor with $r=1/2$, $\beta=0.1$ and $p=10$ from the limit state with $p=-10$, $\tau=\infty$; (right) the hysteresis loops of the same memristor when a time-dependent as a square wave with a large period (>0.2 in dimensionless time units) current is applied to it for $\alpha=\pi/6$ in normalized coordinates resistance--passed charge (top) and current--voltage (bottom), the dots indicate the states corresponding to the profiles shown on the left.}
\end{figure}

For $\beta=0$, the equation~\eqref{eq:kinetics} admits an analytical solution~\cite{boylo2020}. In other cases, its analytical solution is unknown. For many specific values of the parameters, this equation is easy to solve numerically. An example of such a numerical solution, demonstrating the evolution of the vacancy distribution in the memristor when it switches between a stationary (``off'') state at a negative current value $p<0$ and a stationary (``on'') state at $p>0$ is shown in Fig.~\ref{fig:profiles}.

As already mentioned, the total resistance of a closed memristor made of material~\eqref{eq:rho} is determined exclusively by surface effects on its contacts, sensitive to the local concentration of vacancies near them. Considering that $0<c<1$, we can represent the resistance phenomenologically as an expansion in powers of $c$
\begin{align}
 R = & \frac{d}{A}\int_0^1 \left(\rho_0 + \frac{\kappa_{1,0} + \kappa_{1,1} c+ O(c^2)}{d}\delta(\xi) + \right. \nonumber \\
 &\left.+\frac{\kappa_{2,0} + \kappa_{2,1} c+ O(c^2)}{d}\delta(\xi - 1) \right)\ud \xi,
\end{align}
where $A$ is the contact surface area, $\kappa_{1,j}$ and $\kappa_{2,j}$ are the expansion coefficients of  surface resistance at the left and right interfaces in powers of $c$, and the Dirac delta function $\delta(\xi)$ is assumed to be left-handed, which formally places it outside the boundaries of the vacancy motion region. The parameter $\beta$ does not directly affect the memristor resistance $R$. Integrating, in the first order in $c$ we obtain $R=R_0+R_1 c(0,\tau) + R_2 c(1,\tau)$, where $R_0=(d\rho_0+\kappa_{1,0}+\kappa_{2,0})/A$, $R_1=\kappa_{1,1}/A$, $R_2=\kappa_{2,1}/A$. Or, by rearranging the terms, the resistance can be expressed through the dimensionless parameter $\sigma$ as $R=R_0+(R_1+R_2)\sigma$, where
\begin{equation}
  \label{eq:sigma}
  \sigma (\tau)= c(0,\tau)\cos^2\alpha + c(1,\tau) \sin^2\alpha
\end{equation}
and a dimensionless parameter of interface asymmetry $0<\alpha=\arctan \sqrt{R_2/R_1}<\pi/2$ was introduced. In order for the memristor resistances in the ``on'' and ``off'' states to differ, the interfaces must be different ($\alpha\ne\pi/4$). Otherwise, the current-voltage characteristic when passing a current meander (Fig. ~\ref{fig:profiles} bottom right) will have the form of a rotated ``table with legs'', when both inclined straight lines merge into a ``tabletop'', and transient processes (corresponding to vacancy migration at steady-state current values in the meander) will form ``legs''.

\section{Equilibrium states}
The full switching kinetics of the memristor under consideration is currently available only numerically. But when a constant current $|p|= const$ passes through the memristor, it is quite simple and represents a relaxation (at the final stage according to the exponential law) to one of the equilibrium (stationary at $\tau\rightarrow\infty$) states. These states can be found analytically. One of them corresponds to positive values $p>0$, when the vacancies gather at the right ($\xi=1$) end of the memristor, the other --- negative, when the vacancies are at its left ($\xi=0$) end. The vacancy distributions $\lim_{\tau\rightarrow\infty}c(\xi,\tau)=c(\xi)$ in these states are related by a simple transformation
\begin{equation}
\label{eq:simsignP}
 p\rightarrow-p, \quad \xi\rightarrow1-\xi.
\end{equation}
This means that it is enough to find one of them.

This is quite easy to do, since for a stationary $c=c(\xi)$ the equation~\eqref{eq:kinetics} can be integrated over $\xi$ and this integral, as follows from \eqref{eq:continuity}, is proportional to the magnitude of the vacancy flux $J$. The integration constant must be chosen so that the flux at the boundaries of the memristor $\xi=0, 1$ is equal to zero. In the stationary case, this automatically means that the flux is zero everywhere: $J=0$. In the limit $a\rightarrow0$ from~\eqref{eq:current} for $c(x)$ we obtain
\begin{equation}
 -c^\prime + p c (1-c)(1+\beta(c-r)) = 0
\end{equation}
This (albeit nonlinear) ordinary differential equation of the first order is easily solved by the separation of variables, which gives
\begin{equation}
 \xi + C_0 = \frac{\beta  \log \frac{c}{1 + \beta(c- r)}}{p [\beta  (r\!-\!1)\!-\!1]
   (\beta  r\!-\!1)}+\frac{2
   \tanh ^{-1}(1\!-\!2 c) }{p [\beta  (r-1)-1]
   },
\end{equation}
where $C_0$ is the integration constant. The value of $C_0$ can be found from the condition for the total number of vacancies in the memristor~\eqref{eq:r}, which allows us to express the dependence $c(\xi)$ parametrically
\begin{subequations}
\label{eq:parametricC}
\begin{align}
 \xi = & \frac{1}{1 -\beta r} \left[s + \frac{  \beta^2 r^2}{1+\beta  (1-r)} + \right.\nonumber\\
 &\left.
  +\frac{\beta \log
   \frac{e^p-1}{\left(e^p-e^{p s}\right) e^{\beta
   p r^2}+\left(e^{p s}-1\right) e^{(\beta +1) p
   r}}}{p (1+\beta  (1-r))} \right],\\
 c = &\frac{1}{1-\frac{e^{-p s} (1 + \beta  (1 - r)) \left(1 - e^{p
   (1 - r) (1 - \beta  r)}\right)}{(1 - \beta  r) \left(1 -e^{-p
   r (1 + \beta  (1 - r))}\right)}},
\end{align}
\end{subequations}
where the parameter $0\le s \le 1$. For $\beta=0$ the stationary profile $c(\xi)$ can be expressed explicitly
\begin{equation}
 \label{eq:cBurg}
 c_\mathrm{B}(p,r,\xi) = \frac{\left(e^{p r}-1\right) e^{p \xi}}{e^{p (r+\xi)}-e^{p
   r}-e^{p \xi}+e^p}
\end{equation}
and exactly coincides with that obtained in~\cite{boylo2020}. For $\beta\ne0$, several cases are possible, which we will analyze in the following sections.

\subsection{Weak nonlinearity: $-1/(1-r)<\beta<1/r$}
The simplest case is when the parameter $\beta$, controlling the cubic nonlinear term in the equation~\eqref{eq:kinetics}, is small. Let it be so small that the specific resistance of the memristor $\rho>0$ for any $0<c<1$. From~\eqref{eq:rho} this is equivalent to the condition $-1/(1-r)<\beta<1/r$. That is, for each value of the memristor fill factor $r$, the parameter $\beta$ in the weakly nonlinear regime under consideration can take (positive and negative) values from a certain interval around $\beta=0$.

\begin{figure*}[htp]
\begin{center}
\includegraphics[width=1.0\textwidth]{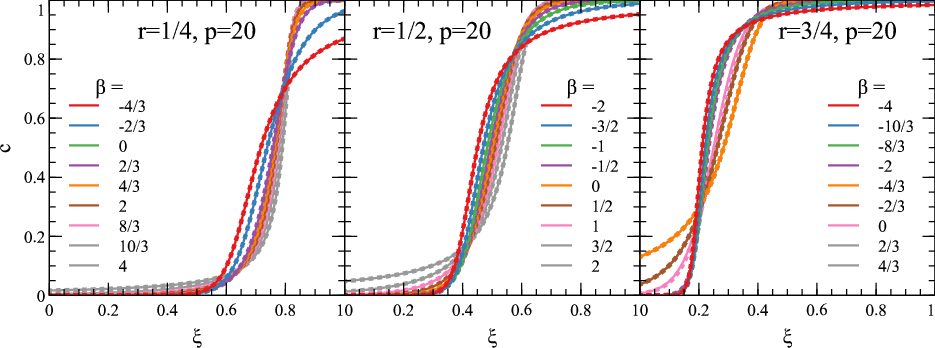}
\vspace{-1.0cm}
\end{center}
\caption{\label{fig:profilesBeta} Distributions of vacancy concentration in equilibrium states at $p=20$ and different memristor fillings $r=1/4$, $1/2$, $3/4$ for different $\beta$ values from $-1/r$ to $1/r$; lines --- analytical expression~\eqref{eq:parametricC}, dots --- numerical calculation.}
\end{figure*}

In Fig.~\ref{fig:profilesBeta}
examples of equilibrium distributions of vacancies in a memristor are given for some (sufficiently large) current $p=20$ and several values of the fill factor $r$ over the entire range of $\beta$ values. Note that the vacancy distributions on the leftmost and rightmost graphs of Fig.~\ref{fig:profilesBeta} are related by successive applications of the similarity transformations~\eqref{eq:betasign} and~\eqref{eq:simsignP}. Therefore, generally speaking, it is sufficient to consider the case $0<r<1/2$.

It can be said that in the weakly nonlinear regime, the influence of the parameter $\beta$ on the equilibrium distributions of vacancies is primarily quantitative. Moreover, from the point of view of achieving the greatest difference between the ``on'' and ``off'' resistances of the memristor, the effect of a non-zero value of $\beta$ is rather negative.

Indeed, for the maximum difference in resistance $\sigma$ from~\eqref{eq:sigma} between the equilibrium states with opposite signs of $p$, it is necessary that the difference in the concentration of vacancies at the opposite ends of the memristor be maximum. From Fig.~\ref{fig:profilesBeta} it is evident that already at $p=20$ this difference at $\beta=0$ almost reaches its maximum value equal to $1$. At the same time, with an increase in $|\beta|$, the concentration of vacancies at the left and right ends of the memristor achieved with the same current $p=20$ begins to differ (and quite significantly) from $0$ and $1$, respectively. Specifically, for the limiting values of $\beta$, the following estimates of the maximum (at the left end of the memristor) and minimum (at its right end) concentration of vacancies can be obtained
\begin{align}
 \max_\beta c(0) & = \frac{\left(e^p-1\right) r}{e^p (r + p (1\!-\!r))-r} \approx\left.
 \frac{r}{r + p (1\!-\!r)}
 \right|_{p\gg1}, \\
 \min_\beta c(1) & = \frac{e^p p r}{e^p (1\!+\!(p\!-\!1) r)\!+\!r\!-\!1}
 \approx \left.
 \frac{p r}{1\!+\!(p\!-\!1) r}
 \right|_{p\gg1}.
\end{align}
It is clear that for $\lim_{p\rightarrow\infty}\max_\beta c(0)=0$ and $\lim_{p\rightarrow\infty}\min_\beta c(1)=1$, but if in the case of $\beta=0$ this limit is reached exponentially at finite values of $p\gg 1$, then for $\beta\ne0$ this happens much more slowly. That is, the resistance of a weakly nonlinear memristor can always be saturated with sufficiently large current, but it is harder to do in memristors with larger values of $|\beta|$.

\subsection{Memristor with phase transition: $r-1<1/\beta<r$}

For large  $|\beta|$ it is possible that, at a certain concentration of vacancies
\begin{equation}
0< c_\mathrm{lim} = \frac{r\beta - 1}{\beta} < 1,
\end{equation}
the specific resistance $\rho$ inside the memristor, determined by the formula~\eqref{eq:rho}, becomes $0$. We will identify this feature with the presence of a phase transition (say, a metal-insulator transition) in the memristor material, when its specific resistance, upon reaching a certain concentration of vacancies, becomes zero (or close to zero) and (practically) stops changing with further changes in $c$.

Metal-insulator transitions are usually observed as a sharp change in electrical resistance by several orders of magnitude when the temperature of the material crosses a certain critical value. But it is known that this critical temperature strongly depends on the concentration and location of oxygen and other vacancies in many materials~\cite{Zhang2012,Wang2016,Chen2016,McGahay2020}. This can be expected {\it a priori}, since the critical temperature of the metal-insulator transition is determined by the electronic structure, which is strongly influenced by charged vacancies. Therefore, at a temperature near the critical one, the metal-insulator transition can also be triggered by a change in the concentration of vacancies. Yes, a linear law with a break may seem to be a rather rough description of the behavior of resistance near the phase transition, but for a sufficiently small neighborhood (in absolute vacancy concentration) of the transition point, the linear approximation is well justified and, at the very least, can give a good qualitative description. In practice, to create memristors operating in a highly nonlinear mode, a fairly precise (one-time for the closed memristors under consideration) fine-tuning of the total number of vacancies and temperature stabilization is required.

Of course, there is no superconductivity in the memristor (and certainly no negative resistance) in this case. In a memristor with phase transition, the concentration of vacancies can only approach the limiting value $c_\mathrm{lim}$, but can never reach it. The reason is that (at a given fixed current through the memristor) a local decrease in resistance reduces the value of the local electric field driving the vacancies. At the same time, the effect of temperature fluctuations, which tend to equalize the local concentration of vacancies at neighboring points, does not weaken. As a result, even if the current is arbitrarily large, the concentration of vacancies is not able to cross the value $c_\mathrm{lim}$. Thus, in a phase transition memristor the values of concentration are no longer in the range $0 <c <1$, but in the reduced range $c_\mathrm{lim}<c<1$ for large positive $\beta>0$ or in the range $0< c < c_\mathrm{lim}$ for (large in absolute value) negative $\beta<0$. The specific resistance remains positive $\rho>0$ in all cases.

\begin{figure}[htp]
\begin{center}
\includegraphics[width=1.0\columnwidth]{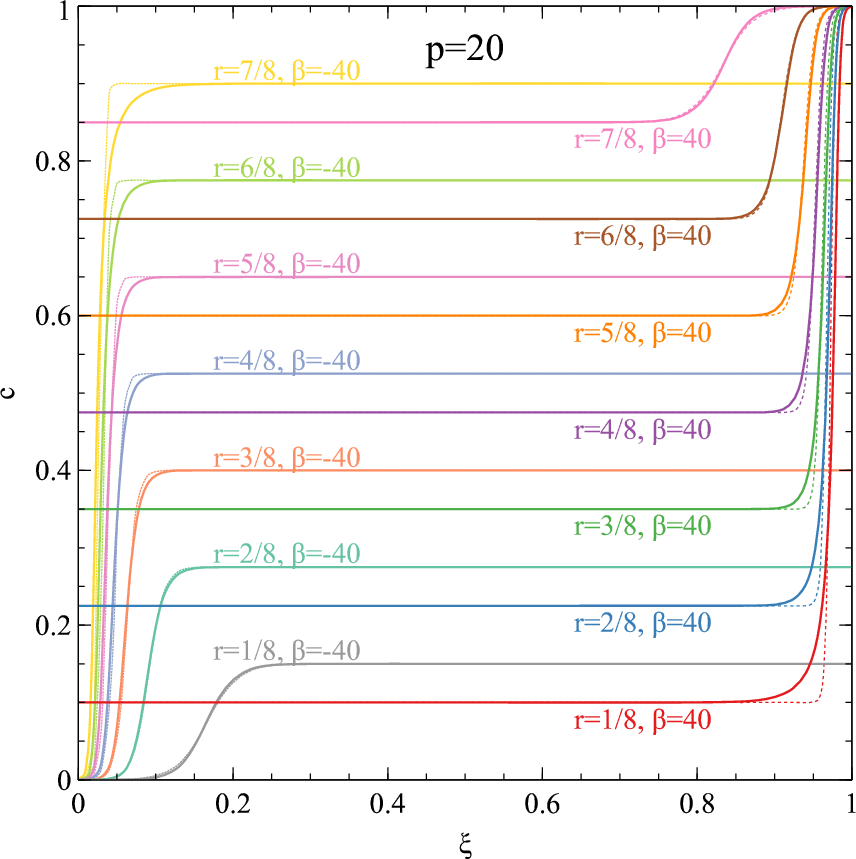}
\vspace{-1cm}
\end{center}
\caption{\label{fig:profilesPhTr} Distributions of vacancy concentration in equilibrium states of a memristor with a phase transition with $p=20$ and $\beta=\pm 40$ for different memristor fillings; lines are the exact analytical expression~\eqref{eq:parametricC}, dots show the approximation~\eqref{eq:burgersAppr}.}
\end{figure}
Examples of equilibrium distributions of vacancies~\eqref{eq:parametricC} in a phase-change memristor are shown for two values of $\beta$ in Fig.~\ref{fig:profilesPhTr} over the entire range of memristor fill factors $r$. In principle, it is sufficient to show only one set of these curves -- either for positive or for negative (with the same modulus) values of $\beta$. The second set can be obtained by successive application of transformations~\eqref{eq:betasign}, \eqref{eq:simsignP}, which boils down to rotating the entire graph by $180^\circ$ around the central point $\xi=1/2$, $c=1/2$ and redesignating the fill factors $r\leftrightarrow (1-r)$.

It can also be noted that with strong nonlinearity, the concentration can be ``clamped'' in a very narrow range of values. For $\beta>0$ this occurs when the memristor is almost completely filled $r\approx 1$, and for $\beta<0$ when $r\approx0$ is almost empty. In this situation, one of the factors in the vacancy current~\eqref{eq:current} $J \propto c \cdot ( 1 - c)$ ceases to have a significant impact on the kinetics and can be replaced by its average value. Then, in the strongly nonlinear case $|\beta|\gg1$, the model~\eqref{eq:kinetics} maps to the rescaled Burgers ($\beta=0$) model:
\begin{align}
 \label{eq:burgersAppr}
 c(\xi,\tau) & \approx c_\mathrm{lim} c_\mathrm{B}(- p r \beta, r/c_\mathrm{lim}, \xi,\tau),\ \text{at}\ \beta<0.
\end{align}
A similar expression for $\beta>0$ can be obtained by the transformation~\eqref{eq:betasign}. Here
$c_B (p,r,\xi,\tau)$ is the solution of the equation~\eqref{eq:kinetics} for $\beta=0$, and $c_\mathrm{B}(p,r,\xi)=\lim_{\tau\rightarrow\infty}c_\mathrm{B}(p,r,\xi,\tau)$ used in plotting the graphs is given by the equation~\eqref{eq:cBurg}. Let us note that this mapping is valid both for stationary states and for transient processes in the memristor.

The equilibrium states corresponding to~\eqref{eq:burgersAppr} are depicted in Fig.~\ref{fig:profilesPhTr} by dotted lines. It can be noted that already at $r\le1/4$ for $\beta=-40$ (and at $r\ge 3/4$ for $\beta=40$) these distributions are visually indistinguishable from the exact ones~\eqref{eq:parametricC}, shown by solid lines. An important advantage of the mapping~\eqref{eq:burgersAppr} is the availability of exact solutions not only for the equilibrium vacancy distributions, but also for time-dependent transient~\cite{boylo2020} and periodic~\cite{BM2024-en} processes. These solutions automatically become applicable for modeling such ``pressed to the phase transition'' strongly nonlinear memristors.

It can be noted that with increasing $|\beta|$, most of the memristor turns out to be filled with vacancies almost at the level of the limiting concentration $c_\mathrm{lim}$, i.e. most of it assumes the state with minimal value of specific resistance. In the considered case of a linear dependence of the resistivity on the vacancy concentration~\eqref{eq:rho} this does not affect the resistance between contacts $\xi=0$ and $\xi=1$. However, if a third contact is added at some intermediate distance $0<\xi<1$ --- the resistance between it and the other two contacts will strongly depend on the state of the memristor. Such a three-terminal device called a memistor~\cite{Xia2011} has already been proposed and implemented~\cite{Kaeriyama2005,Wang2013}, in the context of neuromorphic computing as well~\cite{BP2017}. We hope that the simple theoretical model presented here will help developing intuition, useful for designing and optimizing such devices.

\section{Conclusions}

A model of a memristor made of a material with a linear dependence of its specific resistance on the concentration of oxygen vacancies (or other charged mobile defects) is considered. Its kinetics is determined by a nonlinear partial differential equation, which reduces to the famous Burgers equation in the special case when the dependence of the specific resistance on the vacancy concentration can be neglected. Exact analytical expressions are obtained for the vacancy distributions in the equilibrium states of such a memristor --- spatial distributions of vacancies created by long-term flow of electric current at a constant temperature. In the case of a weak (but still not negligibly small) dependence of the specific resistance on the vacancy concentration, this dependence makes it difficult to saturate the memristor's resistance. Interestingly, the considered simple model applies to memristors, whose material undergoes (depending on the defect concentration) a metal-insulator phase transition. The presence of the transition narrows the permissible range of mobile vacancy concentrations. If the range becomes sufficiently narrow, the switching kinetics of such memristors (and three-terminal memistors) can also be approximately mapped to an exactly solvable Burgers equation with full account of the linear dependence of their resistivity on the vacancy concentration.


\begin{thebibliography}{20}%
\makeatletter
\providecommand \@ifxundefined [1]{%
 \@ifx{#1\undefined}
}%
\providecommand \@ifnum [1]{%
 \ifnum #1\expandafter \@firstoftwo
 \else \expandafter \@secondoftwo
 \fi
}%
\providecommand \@ifx [1]{%
 \ifx #1\expandafter \@firstoftwo
 \else \expandafter \@secondoftwo
 \fi
}%
\providecommand \natexlab [1]{#1}%
\providecommand \enquote  [1]{``#1''}%
\providecommand \bibnamefont  [1]{#1}%
\providecommand \bibfnamefont [1]{#1}%
\providecommand \citenamefont [1]{#1}%
\providecommand \href@noop [0]{\@secondoftwo}%
\providecommand \href [0]{\begingroup \@sanitize@url \@href}%
\providecommand \@href[1]{\@@startlink{#1}\@@href}%
\providecommand \@@href[1]{\endgroup#1\@@endlink}%
\providecommand \@sanitize@url [0]{\catcode `\\12\catcode `\$12\catcode
  `\&12\catcode `\#12\catcode `\^12\catcode `\_12\catcode `\%12\relax}%
\providecommand \@@startlink[1]{}%
\providecommand \@@endlink[0]{}%
\providecommand \url  [0]{\begingroup\@sanitize@url \@url }%
\providecommand \@url [1]{\endgroup\@href {#1}{\urlprefix }}%
\providecommand \urlprefix  [0]{URL }%
\providecommand \Eprint [0]{\href }%
\providecommand \doibase [0]{https://doi.org/}%
\providecommand \selectlanguage [0]{\@gobble}%
\providecommand \bibinfo  [0]{\@secondoftwo}%
\providecommand \bibfield  [0]{\@secondoftwo}%
\providecommand \translation [1]{[#1]}%
\providecommand \BibitemOpen [0]{}%
\providecommand \bibitemStop [0]{}%
\providecommand \bibitemNoStop [0]{.\EOS\space}%
\providecommand \EOS [0]{\spacefactor3000\relax}%
\providecommand \BibitemShut  [1]{\csname bibitem#1\endcsname}%
\let\auto@bib@innerbib\@empty
\bibitem [{\citenamefont {{Chua}}(1971)}]{Chua1971}%
  \BibitemOpen
  \bibfield  {author} {\bibinfo {author} {\bibfnamefont {L.}~\bibnamefont
  {{Chua}}},\ }\bibfield  {title} {\bibinfo {title} {Memristor-the missing
  circuit element},\ }\href {https://doi.org/10.1109/TCT.1971.1083337}
  {\bibfield  {journal} {\bibinfo  {journal} {IEEE Trans. Circuit Theory}\
  }\textbf {\bibinfo {volume} {18}},\ \bibinfo {pages} {507–519} (\bibinfo
  {year} {1971})}\BibitemShut {NoStop}%
\bibitem [{\citenamefont {Chua}\ and\ \citenamefont {Kang}(1976)}]{Chua1976}%
  \BibitemOpen
  \bibfield  {author} {\bibinfo {author} {\bibfnamefont {L.}~\bibnamefont
  {Chua}}\ and\ \bibinfo {author} {\bibfnamefont {S.~M.}\ \bibnamefont
  {Kang}},\ }\bibfield  {title} {\bibinfo {title} {Memristive devices and
  systems},\ }\href {https://doi.org/10.1109/PROC.1976.10092} {\bibfield
  {journal} {\bibinfo  {journal} {Proceedings of the IEEE}\ }\textbf {\bibinfo
  {volume} {64}},\ \bibinfo {pages} {209–223} (\bibinfo {year}
  {1976})}\BibitemShut {NoStop}%
\bibitem [{\citenamefont {Strukov}\ \emph {et~al.}(2008)\citenamefont
  {Strukov}, \citenamefont {Snider}, \citenamefont {Stewart},\ and\
  \citenamefont {Williams}}]{strukov2008}%
  \BibitemOpen
  \bibfield  {author} {\bibinfo {author} {\bibfnamefont {D.~B.}\ \bibnamefont
  {Strukov}}, \bibinfo {author} {\bibfnamefont {G.~S.}\ \bibnamefont {Snider}},
  \bibinfo {author} {\bibfnamefont {D.~R.}\ \bibnamefont {Stewart}},\ and\
  \bibinfo {author} {\bibfnamefont {R.~S.}\ \bibnamefont {Williams}},\
  }\bibfield  {title} {\bibinfo {title} {{The missing memristor found}},\
  }\href {https://doi.org/10.1038/nature06932} {\bibfield  {journal} {\bibinfo
  {journal} {Nature}\ }\textbf {\bibinfo {volume} {453}},\ \bibinfo {pages}
  {80–83} (\bibinfo {year} {2008})}\BibitemShut {NoStop}%
\bibitem [{\citenamefont {{Waser Rainer}}\ and\ \citenamefont {{Aono
  Masakazu}}(2007)}]{Waser2007}%
  \BibitemOpen
  \bibfield  {author} {\bibinfo {author} {\bibnamefont {{Waser Rainer}}}\ and\
  \bibinfo {author} {\bibnamefont {{Aono Masakazu}}},\ }\bibfield  {title}
  {\bibinfo {title} {{Nanoionics-based resistive switching memories}},\ }\href
  {https://doi.org/10.1038/nmat2023} {\bibfield  {journal} {\bibinfo  {journal}
  {Nat. Mater.}\ }\textbf {\bibinfo {volume} {6}},\ \bibinfo {pages}
  {833–840} (\bibinfo {year} {2007})},\ \bibinfo {note}
  {10.1038/nmat2023}\BibitemShut {NoStop}%
\bibitem [{\citenamefont {Sawa}(2008)}]{Sawa2008}%
  \BibitemOpen
  \bibfield  {author} {\bibinfo {author} {\bibfnamefont {A.}~\bibnamefont
  {Sawa}},\ }\bibfield  {title} {\bibinfo {title} {{Resistive switching in
  transition metal oxides}},\ }\href
  {https://doi.org/10.1016/S1369-7021(08)70119-6} {\bibfield  {journal}
  {\bibinfo  {journal} {Mater. Today}\ }\textbf {\bibinfo {volume} {11}},\
  \bibinfo {pages} {28–36} (\bibinfo {year} {2008})}\BibitemShut {NoStop}%
\bibitem [{\citenamefont {{Bryant B.}}\ \emph {et~al.}(2011)\citenamefont
  {{Bryant B.}}, \citenamefont {{Renner Ch.}}, \citenamefont {{Tokunaga Y.}},
  \citenamefont {{Tokura Y.}},\ and\ \citenamefont {{Aeppli G.}}}]{Bryant2011}%
  \BibitemOpen
  \bibfield  {author} {\bibinfo {author} {\bibnamefont {{Bryant B.}}}, \bibinfo
  {author} {\bibnamefont {{Renner Ch.}}}, \bibinfo {author} {\bibnamefont
  {{Tokunaga Y.}}}, \bibinfo {author} {\bibnamefont {{Tokura Y.}}},\ and\
  \bibinfo {author} {\bibnamefont {{Aeppli G.}}},\ }\bibfield  {title}
  {\bibinfo {title} {{Imaging oxygen defects and their motion at a manganite
  surface}},\ }\href {https://doi.org/10.1038/ncomms1219} {\bibfield  {journal}
  {\bibinfo  {journal} {Nat. Commun.}\ }\textbf {\bibinfo {volume} {2}},\
  \bibinfo {pages} {212} (\bibinfo {year} {2011})},\ \bibinfo {note}
  {10.1038/ncomms1219}\BibitemShut {NoStop}%
\bibitem [{\citenamefont {{Yao Lide}}\ \emph {et~al.}(2017)\citenamefont {{Yao
  Lide}}, \citenamefont {{Inkinen Sampo}},\ and\ \citenamefont
  {Sebastiaan}}]{Yao2017}%
  \BibitemOpen
  \bibfield  {author} {\bibinfo {author} {\bibnamefont {{Yao Lide}}}, \bibinfo
  {author} {\bibnamefont {{Inkinen Sampo}}},\ and\ \bibinfo {author}
  {\bibfnamefont {v.~D.}\ \bibnamefont {Sebastiaan}},\ }\bibfield  {title}
  {\bibinfo {title} {{Direct observation of oxygen vacancy-driven structural
  and resistive phase transitions in La2/3Sr1/3MnO3}},\ }\href
  {https://doi.org/10.1038/ncomms14544} {\bibfield  {journal} {\bibinfo
  {journal} {Nat. Commun.}\ }\textbf {\bibinfo {volume} {8}},\ \bibinfo {pages}
  {14544} (\bibinfo {year} {2017})}\BibitemShut {NoStop}%
\bibitem [{\citenamefont {Agudov}\ \emph {et~al.}(2020)\citenamefont {Agudov},
  \citenamefont {Safonov}, \citenamefont {Krichigin}, \citenamefont
  {Kharcheva}, \citenamefont {Dubkov}, \citenamefont {Valenti}, \citenamefont
  {Guseinov}, \citenamefont {Belov}, \citenamefont {Mikhaylov}, \citenamefont
  {Carollo},\ and\ \citenamefont {Spagnolo}}]{agudov2020}%
  \BibitemOpen
  \bibfield  {author} {\bibinfo {author} {\bibfnamefont {N.~V.}\ \bibnamefont
  {Agudov}}, \bibinfo {author} {\bibfnamefont {A.~V.}\ \bibnamefont {Safonov}},
  \bibinfo {author} {\bibfnamefont {A.~V.}\ \bibnamefont {Krichigin}}, \bibinfo
  {author} {\bibfnamefont {A.~A.}\ \bibnamefont {Kharcheva}}, \bibinfo {author}
  {\bibfnamefont {A.~A.}\ \bibnamefont {Dubkov}}, \bibinfo {author}
  {\bibfnamefont {D.}~\bibnamefont {Valenti}}, \bibinfo {author} {\bibfnamefont
  {D.~V.}\ \bibnamefont {Guseinov}}, \bibinfo {author} {\bibfnamefont {A.~I.}\
  \bibnamefont {Belov}}, \bibinfo {author} {\bibfnamefont {A.~N.}\ \bibnamefont
  {Mikhaylov}}, \bibinfo {author} {\bibfnamefont {A.}~\bibnamefont {Carollo}},\
  and\ \bibinfo {author} {\bibfnamefont {B.}~\bibnamefont {Spagnolo}},\
  }\bibfield  {title} {\bibinfo {title} {Nonstationary distributions and
  relaxation times in a stochastic model of memristor},\ }\href
  {https://doi.org/10.1088/1742-5468/ab684a} {\bibfield  {journal} {\bibinfo
  {journal} {J. Stat. Mech.}\ }\textbf {\bibinfo {volume} {2020}},\ \bibinfo
  {pages} {024003} (\bibinfo {year} {2020})}\BibitemShut {NoStop}%
\bibitem [{\citenamefont {Agudov}\ \emph {et~al.}(2021)\citenamefont {Agudov},
  \citenamefont {Dubkov}, \citenamefont {Safonov}, \citenamefont {Krichigin},
  \citenamefont {Kharcheva}, \citenamefont {Guseinov}, \citenamefont
  {Koryazhkina}, \citenamefont {Novikov}, \citenamefont {Shishmakova},
  \citenamefont {Antonov}, \citenamefont {Carollo},\ and\ \citenamefont
  {Spagnolo}}]{agudov2021}%
  \BibitemOpen
  \bibfield  {author} {\bibinfo {author} {\bibfnamefont {N.}~\bibnamefont
  {Agudov}}, \bibinfo {author} {\bibfnamefont {A.}~\bibnamefont {Dubkov}},
  \bibinfo {author} {\bibfnamefont {A.}~\bibnamefont {Safonov}}, \bibinfo
  {author} {\bibfnamefont {A.}~\bibnamefont {Krichigin}}, \bibinfo {author}
  {\bibfnamefont {A.}~\bibnamefont {Kharcheva}}, \bibinfo {author}
  {\bibfnamefont {D.}~\bibnamefont {Guseinov}}, \bibinfo {author}
  {\bibfnamefont {M.}~\bibnamefont {Koryazhkina}}, \bibinfo {author}
  {\bibfnamefont {A.}~\bibnamefont {Novikov}}, \bibinfo {author} {\bibfnamefont
  {V.}~\bibnamefont {Shishmakova}}, \bibinfo {author} {\bibfnamefont
  {I.}~\bibnamefont {Antonov}}, \bibinfo {author} {\bibfnamefont
  {A.}~\bibnamefont {Carollo}},\ and\ \bibinfo {author} {\bibfnamefont
  {B.}~\bibnamefont {Spagnolo}},\ }\bibfield  {title} {\bibinfo {title}
  {Stochastic model of memristor based on the length of conductive region},\
  }\href {https://doi.org/10.1016/j.chaos.2021.111131} {\bibfield  {journal}
  {\bibinfo  {journal} {Chaos, Solitons Fractals}\ }\textbf {\bibinfo {volume}
  {150}},\ \bibinfo {pages} {111131} (\bibinfo {year} {2021})}\BibitemShut
  {NoStop}%
\bibitem [{\citenamefont {Roldán}\ \emph {et~al.}(2023)\citenamefont
  {Roldán}, \citenamefont {Miranda}, \citenamefont {Maldonado}, \citenamefont
  {Mikhaylov}, \citenamefont {Agudov}, \citenamefont {Dubkov}, \citenamefont
  {Koryazhkina}, \citenamefont {González}, \citenamefont {Villena},
  \citenamefont {Poblador}, \citenamefont {Saludes-Tapia}, \citenamefont
  {Picos}, \citenamefont {Jiménez-Molinos}, \citenamefont {Stavrinides},
  \citenamefont {Salvador}, \citenamefont {Alonso}, \citenamefont {Campabadal},
  \citenamefont {Spagnolo}, \citenamefont {Lanza},\ and\ \citenamefont
  {Chua}}]{roldan2023}%
  \BibitemOpen
  \bibfield  {author} {\bibinfo {author} {\bibfnamefont {J.~B.}\ \bibnamefont
  {Roldán}}, \bibinfo {author} {\bibfnamefont {E.}~\bibnamefont {Miranda}},
  \bibinfo {author} {\bibfnamefont {D.}~\bibnamefont {Maldonado}}, \bibinfo
  {author} {\bibfnamefont {A.~N.}\ \bibnamefont {Mikhaylov}}, \bibinfo {author}
  {\bibfnamefont {N.~V.}\ \bibnamefont {Agudov}}, \bibinfo {author}
  {\bibfnamefont {A.~A.}\ \bibnamefont {Dubkov}}, \bibinfo {author}
  {\bibfnamefont {M.~N.}\ \bibnamefont {Koryazhkina}}, \bibinfo {author}
  {\bibfnamefont {M.~B.}\ \bibnamefont {González}}, \bibinfo {author}
  {\bibfnamefont {M.~A.}\ \bibnamefont {Villena}}, \bibinfo {author}
  {\bibfnamefont {S.}~\bibnamefont {Poblador}}, \bibinfo {author}
  {\bibfnamefont {M.}~\bibnamefont {Saludes-Tapia}}, \bibinfo {author}
  {\bibfnamefont {R.}~\bibnamefont {Picos}}, \bibinfo {author} {\bibfnamefont
  {F.}~\bibnamefont {Jiménez-Molinos}}, \bibinfo {author} {\bibfnamefont
  {S.~G.}\ \bibnamefont {Stavrinides}}, \bibinfo {author} {\bibfnamefont
  {E.}~\bibnamefont {Salvador}}, \bibinfo {author} {\bibfnamefont {F.~J.}\
  \bibnamefont {Alonso}}, \bibinfo {author} {\bibfnamefont {F.}~\bibnamefont
  {Campabadal}}, \bibinfo {author} {\bibfnamefont {B.}~\bibnamefont
  {Spagnolo}}, \bibinfo {author} {\bibfnamefont {M.}~\bibnamefont {Lanza}},\
  and\ \bibinfo {author} {\bibfnamefont {L.~O.}\ \bibnamefont {Chua}},\
  }\bibfield  {title} {\bibinfo {title} {Variability in resistive memories},\
  }\href {https://doi.org/10.1002/aisy.202200338} {\bibfield  {journal}
  {\bibinfo  {journal} {Adv. Intell. Syst.}\ }\textbf {\bibinfo {volume} {5}},\
  \bibinfo {pages} {2200338} (\bibinfo {year} {2023})}\BibitemShut {NoStop}%
\bibitem [{\citenamefont {Boylo}\ and\ \citenamefont
  {Metlov}(2021)}]{boylo2020}%
  \BibitemOpen
  \bibfield  {author} {\bibinfo {author} {\bibfnamefont {I.~V.}\ \bibnamefont
  {Boylo}}\ and\ \bibinfo {author} {\bibfnamefont {K.~L.}\ \bibnamefont
  {Metlov}},\ }\bibfield  {title} {\bibinfo {title} {{Nonlinear effects in
  memristors with mobile vacancies}},\ }\href@noop {} {\bibfield  {journal}
  {\bibinfo  {journal} {R. Soc. Open Sci.}\ }\textbf {\bibinfo {volume} {8}},\
  \bibinfo {pages} {210677} (\bibinfo {year} {2021})},\ \Eprint
  {https://arxiv.org/abs/2004.04465} {arXiv:2004.04465 [cond-mat.mes-hall]}
  \BibitemShut {NoStop}%
\bibitem [{\citenamefont {Boylo}\ and\ \citenamefont
  {Metlov}(2024)}]{BM2024-en}%
  \BibitemOpen
  \bibfield  {author} {\bibinfo {author} {\bibfnamefont {I.}~\bibnamefont
  {Boylo}}\ and\ \bibinfo {author} {\bibfnamefont {K.}~\bibnamefont {Metlov}},\
  }\bibfield  {title} {\bibinfo {title} {Frequency dependence of vacancy
  movement hysteresis in a closed memristor based on an exactly solvable model
  of controlled nonlinear diffusion},\ }\href
  {https://journals.rcsi.science/0044-4510/article/view/274800} {\bibfield
  {journal} {\bibinfo  {journal} {JETP}\ }\textbf {\bibinfo {volume} {166}},\
  \bibinfo {pages} {858–867} (\bibinfo {year} {2024})}\BibitemShut {NoStop}%
\bibitem [{\citenamefont {Zhang}\ \emph {et~al.}(2012)\citenamefont {Zhang},
  \citenamefont {Thiess}, \citenamefont {Zalden}, \citenamefont {Zeller},
  \citenamefont {Dederichs}, \citenamefont {Raty}, \citenamefont {Wuttig},
  \citenamefont {Blügel},\ and\ \citenamefont {Mazzarello}}]{Zhang2012}%
  \BibitemOpen
  \bibfield  {author} {\bibinfo {author} {\bibfnamefont {W.}~\bibnamefont
  {Zhang}}, \bibinfo {author} {\bibfnamefont {A.}~\bibnamefont {Thiess}},
  \bibinfo {author} {\bibfnamefont {P.}~\bibnamefont {Zalden}}, \bibinfo
  {author} {\bibfnamefont {R.}~\bibnamefont {Zeller}}, \bibinfo {author}
  {\bibfnamefont {P.~H.}\ \bibnamefont {Dederichs}}, \bibinfo {author}
  {\bibfnamefont {J.-Y.}\ \bibnamefont {Raty}}, \bibinfo {author}
  {\bibfnamefont {M.}~\bibnamefont {Wuttig}}, \bibinfo {author} {\bibfnamefont
  {S.}~\bibnamefont {Blügel}},\ and\ \bibinfo {author} {\bibfnamefont
  {R.}~\bibnamefont {Mazzarello}},\ }\bibfield  {title} {\bibinfo {title} {Role
  of vacancies in metal–insulator transitions of crystalline phase-change
  materials},\ }\href {https://doi.org/10.1038/nmat3456} {\bibfield  {journal}
  {\bibinfo  {journal} {Nat. Mater.}\ }\textbf {\bibinfo {volume} {11}},\
  \bibinfo {pages} {952–956} (\bibinfo {year} {2012})}\BibitemShut {NoStop}%
\bibitem [{\citenamefont {Wang}\ \emph {et~al.}(2016)\citenamefont {Wang},
  \citenamefont {Dash}, \citenamefont {Chang}, \citenamefont {You},
  \citenamefont {Feng}, \citenamefont {He}, \citenamefont {Jin}, \citenamefont
  {Zhou}, \citenamefont {Ong}, \citenamefont {Ren}, \citenamefont {Wang},
  \citenamefont {Chen},\ and\ \citenamefont {Wang}}]{Wang2016}%
  \BibitemOpen
  \bibfield  {author} {\bibinfo {author} {\bibfnamefont {L.}~\bibnamefont
  {Wang}}, \bibinfo {author} {\bibfnamefont {S.}~\bibnamefont {Dash}}, \bibinfo
  {author} {\bibfnamefont {L.}~\bibnamefont {Chang}}, \bibinfo {author}
  {\bibfnamefont {L.}~\bibnamefont {You}}, \bibinfo {author} {\bibfnamefont
  {Y.}~\bibnamefont {Feng}}, \bibinfo {author} {\bibfnamefont {X.}~\bibnamefont
  {He}}, \bibinfo {author} {\bibfnamefont {K.-j.}\ \bibnamefont {Jin}},
  \bibinfo {author} {\bibfnamefont {Y.}~\bibnamefont {Zhou}}, \bibinfo {author}
  {\bibfnamefont {H.~G.}\ \bibnamefont {Ong}}, \bibinfo {author} {\bibfnamefont
  {P.}~\bibnamefont {Ren}}, \bibinfo {author} {\bibfnamefont {S.}~\bibnamefont
  {Wang}}, \bibinfo {author} {\bibfnamefont {L.}~\bibnamefont {Chen}},\ and\
  \bibinfo {author} {\bibfnamefont {J.}~\bibnamefont {Wang}},\ }\bibfield
  {title} {\bibinfo {title} {Oxygen vacancy induced room-temperature
  metal–insulator transition in nickelate films and its potential application
  in photovoltaics},\ }\href {https://doi.org/10.1021/acsami.6b00650}
  {\bibfield  {journal} {\bibinfo  {journal} {ACS Appl. Mater. Interfaces}\
  }\textbf {\bibinfo {volume} {8}},\ \bibinfo {pages} {9769–9776} (\bibinfo
  {year} {2016})}\BibitemShut {NoStop}%
\bibitem [{\citenamefont {Chen}\ \emph {et~al.}(2016)\citenamefont {Chen},
  \citenamefont {Wang}, \citenamefont {Wan}, \citenamefont {Cui}, \citenamefont
  {Liu}, \citenamefont {Shi}, \citenamefont {Luo},\ and\ \citenamefont
  {Gao}}]{Chen2016}%
  \BibitemOpen
  \bibfield  {author} {\bibinfo {author} {\bibfnamefont {L.}~\bibnamefont
  {Chen}}, \bibinfo {author} {\bibfnamefont {X.}~\bibnamefont {Wang}}, \bibinfo
  {author} {\bibfnamefont {D.}~\bibnamefont {Wan}}, \bibinfo {author}
  {\bibfnamefont {Y.}~\bibnamefont {Cui}}, \bibinfo {author} {\bibfnamefont
  {B.}~\bibnamefont {Liu}}, \bibinfo {author} {\bibfnamefont {S.}~\bibnamefont
  {Shi}}, \bibinfo {author} {\bibfnamefont {H.}~\bibnamefont {Luo}},\ and\
  \bibinfo {author} {\bibfnamefont {Y.}~\bibnamefont {Gao}},\ }\bibfield
  {title} {\bibinfo {title} {Tuning the phase transition temperature{,}
  electrical and optical properties of vo2 by oxygen nonstoichiometry: insights
  from first-principles calculations},\ }\href
  {https://doi.org/10.1039/C6RA09449J} {\bibfield  {journal} {\bibinfo
  {journal} {RSC Adv.}\ }\textbf {\bibinfo {volume} {6}},\ \bibinfo {pages}
  {73070–73082} (\bibinfo {year} {2016})}\BibitemShut {NoStop}%
\bibitem [{\citenamefont {McGahay}\ \emph {et~al.}(2020)\citenamefont
  {McGahay}, \citenamefont {Khare},\ and\ \citenamefont {Gall}}]{McGahay2020}%
  \BibitemOpen
  \bibfield  {author} {\bibinfo {author} {\bibfnamefont {M.~E.}\ \bibnamefont
  {McGahay}}, \bibinfo {author} {\bibfnamefont {S.~V.}\ \bibnamefont {Khare}},\
  and\ \bibinfo {author} {\bibfnamefont {D.}~\bibnamefont {Gall}},\ }\bibfield
  {title} {\bibinfo {title} {Metal-insulator transitions in epitaxial
  rocksalt-structure $cr_{1-x/2}n_{1-x}o_x$ (001)},\ }\bibfield  {journal}
  {\bibinfo  {journal} {Physical Review B}\ }\textbf {\bibinfo {volume}
  {102}},\ \href {https://doi.org/10.1103/physrevb.102.235102}
  {10.1103/physrevb.102.235102} (\bibinfo {year} {2020})\BibitemShut {NoStop}%
\bibitem [{\citenamefont {Xia}\ \emph {et~al.}(2011)\citenamefont {Xia},
  \citenamefont {Pickett}, \citenamefont {Yang}, \citenamefont {Li},
  \citenamefont {Wu}, \citenamefont {Medeiros‐Ribeiro},\ and\ \citenamefont
  {Williams}}]{Xia2011}%
  \BibitemOpen
  \bibfield  {author} {\bibinfo {author} {\bibfnamefont {Q.}~\bibnamefont
  {Xia}}, \bibinfo {author} {\bibfnamefont {M.~D.}\ \bibnamefont {Pickett}},
  \bibinfo {author} {\bibfnamefont {J.~J.}\ \bibnamefont {Yang}}, \bibinfo
  {author} {\bibfnamefont {X.}~\bibnamefont {Li}}, \bibinfo {author}
  {\bibfnamefont {W.}~\bibnamefont {Wu}}, \bibinfo {author} {\bibfnamefont
  {G.}~\bibnamefont {Medeiros‐Ribeiro}},\ and\ \bibinfo {author}
  {\bibfnamefont {R.~S.}\ \bibnamefont {Williams}},\ }\bibfield  {title}
  {\bibinfo {title} {Two‐ and three‐terminal resistive switches:
  Nanometer‐scale memristors and memistors},\ }\href
  {https://doi.org/10.1002/adfm.201100180} {\bibfield  {journal} {\bibinfo
  {journal} {Advanced Functional Materials}\ }\textbf {\bibinfo {volume}
  {21}},\ \bibinfo {pages} {2660–2665} (\bibinfo {year} {2011})}\BibitemShut
  {NoStop}%
\bibitem [{\citenamefont {Kaeriyama}\ \emph {et~al.}(2005)\citenamefont
  {Kaeriyama}, \citenamefont {Sakamoto}, \citenamefont {Sunamura},
  \citenamefont {Mizuno}, \citenamefont {Kawaura}, \citenamefont {Hasegawa},
  \citenamefont {Terabe}, \citenamefont {Nakayama},\ and\ \citenamefont
  {Aono}}]{Kaeriyama2005}%
  \BibitemOpen
  \bibfield  {author} {\bibinfo {author} {\bibfnamefont {S.}~\bibnamefont
  {Kaeriyama}}, \bibinfo {author} {\bibfnamefont {T.}~\bibnamefont {Sakamoto}},
  \bibinfo {author} {\bibfnamefont {H.}~\bibnamefont {Sunamura}}, \bibinfo
  {author} {\bibfnamefont {M.}~\bibnamefont {Mizuno}}, \bibinfo {author}
  {\bibfnamefont {H.}~\bibnamefont {Kawaura}}, \bibinfo {author} {\bibfnamefont
  {T.}~\bibnamefont {Hasegawa}}, \bibinfo {author} {\bibfnamefont
  {K.}~\bibnamefont {Terabe}}, \bibinfo {author} {\bibfnamefont
  {T.}~\bibnamefont {Nakayama}},\ and\ \bibinfo {author} {\bibfnamefont
  {M.}~\bibnamefont {Aono}},\ }\bibfield  {title} {\bibinfo {title} {A
  nonvolatile programmable solid-electrolyte nanometer switch},\ }\href
  {https://doi.org/10.1109/jssc.2004.837244} {\bibfield  {journal} {\bibinfo
  {journal} {IEEE Journal of Solid-State Circuits}\ }\textbf {\bibinfo {volume}
  {40}},\ \bibinfo {pages} {168–176} (\bibinfo {year} {2005})}\BibitemShut
  {NoStop}%
\bibitem [{\citenamefont {Wang}\ \emph {et~al.}(2013)\citenamefont {Wang},
  \citenamefont {Itoh}, \citenamefont {Hasegawa}, \citenamefont {Tsuruoka},
  \citenamefont {Yamaguchi}, \citenamefont {Watanabe}, \citenamefont
  {Hiramoto},\ and\ \citenamefont {Aono}}]{Wang2013}%
  \BibitemOpen
  \bibfield  {author} {\bibinfo {author} {\bibfnamefont {Q.}~\bibnamefont
  {Wang}}, \bibinfo {author} {\bibfnamefont {Y.}~\bibnamefont {Itoh}}, \bibinfo
  {author} {\bibfnamefont {T.}~\bibnamefont {Hasegawa}}, \bibinfo {author}
  {\bibfnamefont {T.}~\bibnamefont {Tsuruoka}}, \bibinfo {author}
  {\bibfnamefont {S.}~\bibnamefont {Yamaguchi}}, \bibinfo {author}
  {\bibfnamefont {S.}~\bibnamefont {Watanabe}}, \bibinfo {author}
  {\bibfnamefont {T.}~\bibnamefont {Hiramoto}},\ and\ \bibinfo {author}
  {\bibfnamefont {M.}~\bibnamefont {Aono}},\ }\bibfield  {title} {\bibinfo
  {title} {Nonvolatile three-terminal operation based on oxygen vacancy drift
  in a pt/ta\_2o\_{5-x}/pt, pt structure},\ }\href
  {https://doi.org/10.1063/1.4811122} {\bibfield  {journal} {\bibinfo
  {journal} {Applied Physics Letters}\ }\textbf {\bibinfo {volume} {102}},\
  \bibinfo {pages} {233508} (\bibinfo {year} {2013})}\BibitemShut {NoStop}%
\bibitem [{\citenamefont {{Balakrishna Pillai}}\ and\ \citenamefont {{De
  Souza}}(2017)}]{BP2017}%
  \BibitemOpen
  \bibfield  {author} {\bibinfo {author} {\bibfnamefont {P.}~\bibnamefont
  {{Balakrishna Pillai}}}\ and\ \bibinfo {author} {\bibfnamefont {M.~M.}\
  \bibnamefont {{De Souza}}},\ }\bibfield  {title} {\bibinfo {title}
  {Nanoionics-based three-terminal synaptic device using zinc oxide},\ }\href
  {https://doi.org/10.1021/acsami.6b13746} {\bibfield  {journal} {\bibinfo
  {journal} {ACS Applied Materials \&amp; Interfaces}\ }\textbf {\bibinfo
  {volume} {9}},\ \bibinfo {pages} {1609–1618} (\bibinfo {year}
  {2017})}\BibitemShut {NoStop}%
\end{thebibliography}
%
\end{document}


\selectlanguage{russian}
\title{Переключение закрытого мемристора на подвижных вакансиях в материале с линейной зависимостью удельного сопротивления от их концентрации}
\author{\firstname{И.~В.} \surname{Бойло}}
\ifarxiv\email[]{boylo@donfti.ru}\fi
\affiliation{Донецкий Физико-Технический Институт им. А.А. Галкина, 283048 г. Донецк, ДНР, Россия}
\author{\firstname{К.~Л.} \surname{Метлов}}
\ifarxiv\email[]{metlov@donfti.ru}\fi
\affiliation{Донецкий Физико-Технический Институт им. А.А. Галкина, 283048 г. Донецк, ДНР, Россия}
\date{\today}
\begin{abstract}
Линейная по концентрации добавка к удельному сопротивлению материала не дает прямого вклада в общее сопротивление мемристора при перераспределении вакансий с сохранением их общего числа. Но сопутствующее локальное изменение удельного сопротивления меняет кинетику дрейфа вакансий под действием приложенного электрического тока. Эти изменения особенно значительны при наличии в материале мемристора фазового перехода металл-изолятор. В работе получено кинетическое уравнение для локальной концентрации вакансий в таком мемристоре, проанализированы точные решения для его стационарных состояний. Показано, что не только в слабо-нелинейном случае, когда зависимостью удельного сопротивления от концентрации вакансий можно пренебречь, но и в сильно-нелинейном мемристоре с фазовым переходом, его кинетику можно свести к классическому точно решаемому уравнению Бюргерса.
\end{abstract}
\maketitle
\section{Введение}
 Мемристор --- предложенный Леоном Чуа~\cite{Chua1971,Chua1976} двухтерминальный элемент электрических цепей, сопротивление которого зависит от суммарного прошедшего через него электрического заряда. Такой элемент был реализован фирмой Hewlett Packard на основе миграции заряженных кислородных вакансий в диоксиде титана~\cite{strukov2008}. Его можно изготовить и на основе других материалов, допускающих миграцию ионов~\cite{Waser2007} --- различных оксидов переходных металлов~\cite{Sawa2008}, манганитов~\cite{Bryant2011,Yao2017}. Основным свойством мемристора является наличие гистерезиса на его вольт-амперной характеристике. Гистерезис возникает потому, что состояние мемристора определяется не мгновенной величиной проходящего через него электрического тока, а зависит ещё и от величины тока в прошлом (т.е. от истории внешних воздействий). В мемристоре на подвижных вакансиях состояние представлено в виде некоторого пространственного распределения концентрации вакансий, а его изменение происходит посредством управляемой внешним электрическим током диффузии. Измеряемое же сопротивление мемристора электрическому току отражает мгновенное распределение вакансий.

 В литературе изменение состояния таких мемристоров моделируется различными кинетическими уравнениями для концентрации вакансий~\cite{agudov2020,agudov2021,roldan2023}, в основном, численно. Если удельное сопротивление материала считать постоянной величиной --- нелинейная кинетика мемристора описывается уравнением Бюргерса. Такой подход допускает точное аналитическое решение как для переходных процессов~\cite{boylo2020} в закрытом мемристоре (с постоянным общим количеством вакансий в нём), так и периодических~\cite{BM2024}, возникающих при пропускании через такой мемристор токового меандра. Изменение сопротивления закрытого мемристора с постоянным удельным сопротивлением материала определяется исключительно интерфейсными эффектами и зависит от концентрацией подвижных вакансий в области подключенных к нему контактов~\cite{boylo2020}.

 Если удельное сопротивление материала мемристора $\rho$ зависит от концентрации вакансий $c$ линейно: $\rho=a + \beta\cdot c$, где $a, \beta=const$ --- переменная составляющая полного сопротивления закрытого мемристора всё так же определяется исключительно интерфейсами. Объёмный вклад в сопротивление зависит только от полного количества вакансий, а значит в закрытом мемристоре не зависит от их (определяемого историей внешних воздействий) пространственного распределения. Но вот кинетика перераспределения вакансий от коэффициента $\beta$ всё же зависит и при определённых условиях эта зависимость способна проявить себя достаточно сильно. Именно эти интересные с нашей точки зрения эффекты детально исследуется в настоящей работе.

 \section{Модель}
 Рассмотрим тонкую пленку толщиной $d$ из материала, содержащего заряженные подвижные вакансии, ограниченную двумя плоскими, непроницаемыми для вакансий, хорошо проводящими контактами. Пусть координата $x$ отсчитывается в направлении, перпендикулярном поверхности пленки $0\le x \le d$. Состояние такого мемристора в момент времени $t$ описывается мгновенной локальной концентрацией мобильных вакансий $C(x,t)$. Причём, их общее количество $\int_0^d C \ud x$ является постоянной, не зависящей от времени, величиной. В силу локального сохранения количества вакансий, их движение подчиняется уравнению непрерывности
 \begin{equation}
  \label{eq:continuity}
 \partial_t C(x,t) + \vec{\nabla}\cdot\vec{J}(x,t) = 0,
 \end{equation}
где $\partial_t$ производная по времени. В одномерном случае ток вакансий $\vec{J}=\{J,0,0\}$, а $\vec{\nabla}\cdot\vec{J}(x,t)=\partial_x J(x,t)$. Рассматривая диффузию и прыжковый термически-активированный транспорт вакансий под действием электрического тока, представим~\cite{boylo2020}
\begin{equation}
 \label{eq:current}
 J = -D\partial_x C + C (1-C/C_\mathrm{max}) \frac{2 D}{a}\sinh\frac{a q \rho I}{\ukB T},
\end{equation}
где $D$ --- коэффициент диффузии, $q$ --- заряд одной вакансии, $I$ --- величина пропускаемого через мемристор тока, $\ukB$ --- постоянная Больцмана, $T$ -- температура, $a$ --- длина прыжка вакансий (расстояние между соседними равновесными позициями вакансий в материале), $\rho$ --- удельное сопротивление материала. Здесь также считается, что локальная концентрация вакансий ограничена некоторой наперёд заданной величиной $C_\mathrm{max}$, являющейся свойством материала~\cite{boylo2020}.

Перейдем к безразмерной координате $\xi=x/d$, безразмерному времени $\tau=t D/d^2$ и введём нормированную концентрацию вакансий $c=C/C_\mathrm{max}$, $0\le c(\xi,\tau) \le 1$. Тогда линейную зависимость удельного сопротивления от концентрации вакансий удобно представить
как
\begin{equation}
 \label{eq:rho}
 \rho = \rho_0 (1 + \beta (c(\xi,\tau) - r)),
\end{equation}
где коэффициент заполнения мемристора вакансиями $r$ ---
\begin{equation}
\label{eq:r}
 0 < r=\int_0^1 c(\xi,\tau) \ud \xi = const < 1,
\end{equation}
а $\rho_0 = \int_0^1 \rho \ud \xi = const$ --- среднее удельное сопротивление материала мемристора. Параметр $\beta$ характеризует силу влияния концентрации вакансий на удельное сопротивление и может быть как положительным, так и отрицательным. Подставляя~\eqref{eq:rho} в~\eqref{eq:current} и~\eqref{eq:continuity}, переходя к пределу непрерывного дрейфа $a\rightarrow 0$ для нормированной концентрации вакансий получим уравнение
\begin{equation}
 \label{eq:kinetics}
 \partial_\tau c + p \left(1\!-\!r\beta\!-\!c \left(2 \!-\!2 (1\!+\!r)\beta\!+\!3\beta c\right)\right)\partial_\xi c - \partial_{\xi\xi}c = 0,
\end{equation}
где параметр $p=2 q d \rho_0 I/(\ukB T)$ играет роль безразмерной эффективной величины проходящего через мемристор тока. В пределе $\beta\rightarrow 0$ это уравнение сводится к классическому уравнению Бюргерса~\cite{boylo2020}, но в общем случае содержит нелинейность более высокого 3-го порядка по $c$. Задача о кинетике рассматриваемого замкнутого мемристора сводится к решению этого уравнения с граничными условиями $J|_{x=0} = J|_{x=d} =0$.

Интересно, что решения  уравнения~\eqref{eq:kinetics} при положительных и отрицательных значениях параметра $\beta$ связаны преобразованием
\begin{equation}
 \label{eq:betasign}
 \beta\rightarrow-\beta,\quad
 p\rightarrow-p,\quad r\rightarrow1-r,\quad c\rightarrow1-c.
\end{equation}
Благодаря ему, достаточно научиться решать~\eqref{eq:kinetics} только при положительных (или только отрицательных) значениях $\beta$, а решения с $\beta$ противоположного знака можно получить потом простым переопределением параметров~\eqref{eq:betasign}. Так же преобразуются и все следствия из уравнения~\eqref{eq:kinetics}.

\begin{figure}
\begin{center}
\includegraphics[width=1.0\columnwidth]{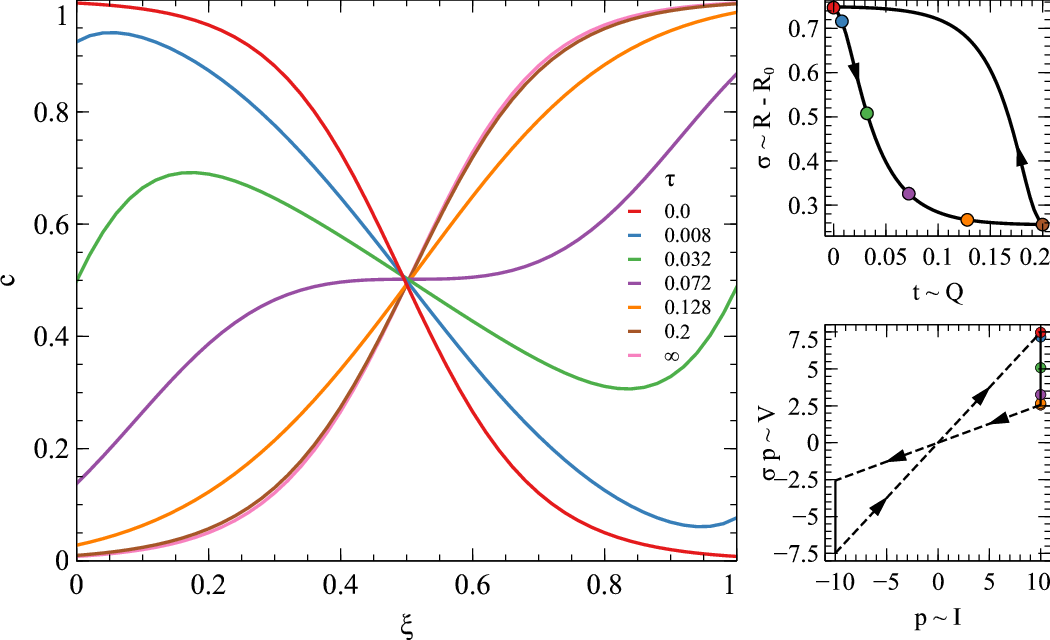}
\vspace{-0.8cm}
\end{center}
\caption{\label{fig:profiles} (слева) Полученные численным решением уравнения~\eqref{eq:kinetics}  профили концентрации $c(\xi,\tau)$ в разные времена  $\tau$ при переключении мемристора с $r=1/2$, $\beta=0.1$ и $p=10$ из равновесного состояния с $p=-10$, $\tau=\infty$; (справа) петли гистерезиса того же мемристора при подаче на него меандра с большим периодом (>0.2 в безразмерных единицах времени) для $\alpha=\pi/6$ в нормированных координатах сопротивление -- прошедший заряд (вверху) и ток--напряжение (внизу), точками показаны состояния, соответствующие профилям, изображённым слева}
\end{figure}

При $\beta=0$ уравнение~\eqref{eq:kinetics} допускает аналитическое решение~\cite{boylo2020}. В остальных случаях, его аналитическое решение неизвестно. Для многих конкретных значений параметров это уравнение несложно решить численно. Пример такого численного решения, демонстрирующий эволюцию распределения вакансий в мемристоре при его переключении между стационарным (``выключенным'') состоянием при отрицательном значении тока $p<0$  и стационарным (``включенным'') состоянием при $p>0$ показан на Рис.~\ref{fig:profiles}.

Как уже было сказано, полное сопротивление закрытого мемристора из материала~\eqref{eq:rho} определяется исключительно поверхностными эффектами на его контактах, чувствительными к локальной концентрации вакансий вблизи них. Учитывая, что $0<c<1$, представим его феноменологически в виде разложения по степеням $c$
\begin{align}
 R = & \frac{d}{A}\int_0^1 \left(\rho_0 + \frac{\kappa_{1,0} + \kappa_{1,1} c+ O(c^2)}{d}\delta(\xi) + \right. \nonumber \\
 &\left.+\frac{\kappa_{2,0} + \kappa_{2,1} c+ O(c^2)}{d}\delta(\xi - 1) \right)\ud \xi,
\end{align}
где $A$ --- площадь поверхности контакта, $\kappa_{1,j}$ и $\kappa_{2,j}$ --- коэффициенты разложения поверхностного сопротивления на левом и правом интерфейсах по степеням $c$, а дельта-функция Дирака $\delta(\xi)$ предполагается левосторонней, что формально выносит её за границы области движения вакансий. Параметр $\beta$ на сопротивление мемристора непосредственно не влияет. Интегрируя, в первом порядке по $c$ получим $R=R_0+R_1 c(0,\tau) + R_2 c(1,\tau)$, где $R_0=(d\rho_0+\kappa_{1,0}+\kappa_{2,0})/A$, $R_1=\kappa_{1,1}/A$, $R_2=\kappa_{2,1}/A$. Или, перегруппировкой членов сопротивление можно выразить через безразмерный параметр $\sigma$ как $R=R_0+(R_1+R_2)\sigma$, где
\begin{equation}
  \label{eq:sigma}
  \sigma (\tau)= c(0,\tau)\cos^2\alpha + c(1,\tau) \sin^2\alpha,
\end{equation}
и введен безразмерный параметр асимметрии интерфейсов $0<\alpha=\arctan \sqrt{R_2/R_1}<\pi/2$. Для того, чтобы сопротивления мемристора во ``включенном'' и ``выключенном'' состояниях отличались, интерфейсы должны быть разными ($\alpha\ne\pi/4$). В противном случае, вольт-амперная характеристика при пропускании меандра тока (Рис. ~\ref{fig:profiles} справа внизу) будет иметь вид повёрнутого ``стола с ножками'' (table with legs), когда обе наклонные прямые линии сольются в ``столешницу'', а переходные процессы (соответствующие миграции вакансий при стационарных значениях тока в меандре) сформируют ``ножки''.

\section{Равновесные состояния}

Полная кинетика переключений рассматриваемого мемристора доступна на данный момент только численно. Но при прохождении через мемристор постоянного тока $|p|= const$, она достаточно простая и представляет собой релаксацию (на конечном этапе по экспоненциальному закону) к одному из равновесных (стационарных при $\tau\rightarrow\infty$) состояний. Эти состояния можно найти аналитически. Одно из них соответствует положительным значениям $p>0$, когда вакансии собираются у правого ($\xi=1$) конца мемристора, другое --- отрицательным, когда вакансии собираются у левого ($\xi=0$) его конца. Распределения вакансий $\lim_{\tau\rightarrow\infty}c(\xi,\tau)=c(\xi)$ в этих состояниях связаны простым преобразованием
\begin{equation}
\label{eq:simsignP}
 p\rightarrow-p, \quad \xi\rightarrow1-\xi.
\end{equation}
Это означает, что достаточно найти одно из них.

Сделать это достаточно просто, поскольку для стационарного $c=c(\xi)$ уравнение~\eqref{eq:kinetics} можно проинтегрировать по $\xi$ и этот интеграл, как следует из \eqref{eq:continuity}, пропорционален величине потока вакансий. Константу интегрирования нужно выбрать так, чтобы поток на границах мемристора $\xi=0,1$ был равен нулю. В стационарном случае это автоматически означает, что поток равен нулю везде: $J=0$. В пределе $a\rightarrow0$ из~\eqref{eq:current} для $c(x)$ получим
\begin{equation}
 -c^\prime + p c (1-c)(1+\beta(c-r)) = 0
\end{equation}
это (пусть и нелинейное) обыкновенное дифференциальное уравнение первого порядка легко решается методом разделения переменных, что даёт
\begin{equation}
 \xi + C_0 = \frac{\beta  \log \frac{c}{1 + \beta(c- r)}}{p [\beta  (r\!-\!1)\!-\!1]
   (\beta  r\!-\!1)}+\frac{2
   \tanh ^{-1}(1\!-\!2 c) }{p [\beta  (r-1)-1]
   },
\end{equation}
где $C_0$ --- постоянная интегрирования. Величину $C_0$ можно найти из условия для полного количества вакансий в мемристоре~\eqref{eq:r} и выразить зависимость $c(\xi)$ параметрически:
\begin{subequations}
\label{eq:parametricC}
\begin{align}
 \xi = & \frac{1}{1 -\beta r} \left[s + \frac{  \beta^2 r^2}{1+\beta  (1-r)} + \right.\nonumber\\
 &\left.
  +\frac{\beta \log
   \frac{e^p-1}{\left(e^p-e^{p s}\right) e^{\beta
   p r^2}+\left(e^{p s}-1\right) e^{(\beta +1) p
   r}}}{p (1+\beta  (1-r))} \right],\\
 c = &\frac{1}{1-\frac{e^{-p s} (1 + \beta  (1 - r)) \left(1 - e^{p
   (1 - r) (1 - \beta  r)}\right)}{(1 - \beta  r) \left(1 -e^{-p
   r (1 + \beta  (1 - r))}\right)}},
\end{align}
\end{subequations}
где параметр $0\le s \le 1$.  При $\beta=0$ стационарный профиль $c(\xi)$ выражается явно
\begin{equation}
 \label{eq:cBurg}
 c_\mathrm{B}(p,r,\xi) = \frac{\left(e^{p r}-1\right) e^{p \xi}}{e^{p (r+\xi)}-e^{p
   r}-e^{p \xi}+e^p}
\end{equation}
и в точности совпадает с полученным в~\cite{boylo2020}. Для $\beta\ne0$ возможны несколько случаев, которые мы проанализируем в следующих разделах.

\subsection{Слабая нелинейность: $-1/(1-r)<\beta<1/r$}

Самый простой случай, когда параметр $\beta$, определяющий наличие кубического нелинейного члена в уравнении~\eqref{eq:kinetics}, мал. Конкретно рассмотрим случай, когда он мал настолько, что удельное сопротивление мемристора $\rho>0$ для любых $0<c<1$. Из~\eqref{eq:rho} это эквивалентно условию $-1/(1-r)<\beta<1/r$. То есть, при каждом значении коэффициента заполнения мемристора $r$ параметр $\beta$ в рассматриваемом слабо-нелинейном режиме может принимать (положительные и отрицательные) значения из некоторого интервала вокруг $\beta=0$.

\begin{figure*}
\begin{center}
\includegraphics[width=1.0\textwidth]{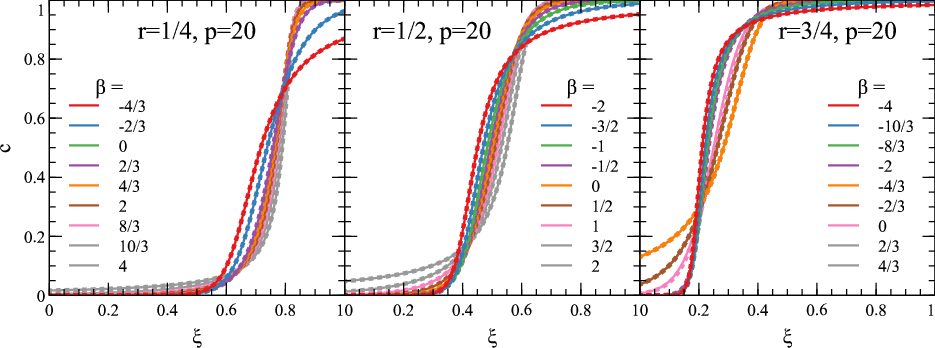}
\vspace{-1cm}
\end{center}
\caption{\label{fig:profilesBeta} Распределения концентрации вакансий в равновесных состояниях при $p=20$ и разных заполнениях мемристора $r=1/4$, $1/2$, $3/4$ для различных значений $\beta$ от $-1/r$ до $1/r$; линии --- аналитическое выражение~\eqref{eq:parametricC}, точки --- численный счёт}
\end{figure*}

На Рис.~\ref{fig:profilesBeta}
приведены примеры равновесных распределений вакансий в мемристоре при некотором (достаточно большом) токе $p=20$ и нескольких значениях коэффициента заполнения $r$ во всём интервале значений $\beta$. Обратите внимание, что распределения вакансий на крайнем левом и крайнем правом графиках Рис.~\ref{fig:profilesBeta} связаны последовательным применением преобразований подобия~\eqref{eq:betasign} и~\eqref{eq:simsignP}. Поэтому, вообще говоря, достаточно рассмотреть случай $0<r<1/2$.

Можно сказать, что в слабо-нелинейном режиме влияние параметра $\beta$ на равновесные распределения вакансий прежде всего количественное. Причём, с точки зрения достижения наибольшей разницы между ``включенным'' и ``выключенным'' сопротивлениями мемристора, эффект от ненулевого значения $\beta$, скорее, отрицательный.

Действительно, для максимальной разницы сопротивления $\sigma$ из~\eqref{eq:sigma} между равновесными состояниями с противоположными знаками $p$ необходимо, чтобы разница в концентрации вакансий на противоположных концах мемристора была максимальной. Из Рис.~\ref{fig:profilesBeta} видно, что уже при $p=20$ эта разность при $\beta=0$ почти достигает своего максимального значения, равного $1$. В то же время, при увеличении $|\beta|$ достигаемая с тем же током $p=20$ концентрация вакансий на левом и правом концах мемристора начинает (и довольно существенно) отличаться от $0$ и $1$, соответственно. Конкретно, для предельных значений $\beta$ можно получить следующие оценки максимальной (на левом конце мемристора) и минимальной (на его правом конце) концентрации вакансий
\begin{align}
 \max_\beta c(0) & = \frac{\left(e^p-1\right) r}{e^p (r + p (1\!-\!r))-r} \approx\left.
 \frac{r}{r + p (1\!-\!r)}
 \right|_{p\gg1}, \\
 \min_\beta c(1) & = \frac{e^p p r}{e^p (1\!+\!(p\!-\!1) r)\!+\!r\!-\!1}
 \approx \left.
 \frac{p r}{1\!+\!(p\!-\!1) r}
 \right|_{p\gg1}.
\end{align}
Понятно, что при $\lim_{p\rightarrow\infty}\max_\beta c(0)=0$ и  $\lim_{p\rightarrow\infty}\min_\beta c(1)=1$, но если в случае $\beta=0$ этот предел при конечных значениях $p\gg 1$ достигается экспоненциально, то при $\beta\ne0$ это происходит гораздо медленнее. То есть, сопротивление слабо-нелинейного мемристора всегда можно насытить достаточно большим током, но чем больше $|\beta|$, тем сделать это сложнее.

\subsection{Мемристор с фазовым переходом: $r-1<1/\beta<r$}

При больших значениях $|\beta|$ возможна ситуация, когда при некоторой концентрации вакансий
\begin{equation}
 0< c_\mathrm{lim} = \frac{r\beta - 1}{\beta} < 1
\end{equation}
удельное сопротивление $\rho$ внутри мемристора, определяемое формулой~\eqref{eq:rho}, обращается в $0$. Мы будем идентифицировать эту особенность с наличием фазового перехода (скажем, перехода металл-изолятор) в материале мемристора, когда его удельное сопротивление при достижении определённой концентрации вакансий обращается в ноль (или становится близким к нулю) и при дальнейшем изменении их концентрации (практически) перестаёт меняться.

Обычно переходы металл-изолятор наблюдают как резкое изменение электрического сопротивления на несколько порядков при изменении температуры материала. Но известно, что критическая температура этого перехода во многих материалах сильно зависит от концентрации и расположения  вакансий~\cite{Zhang2012,Wang2016,Chen2016,McGahay2020}. Этого можно ожидать и {\it a priori}, поскольку критическая температура перехода металл-изолятор определяется электронной структурой, а заряженные вакансии способны эту структуру существенным образом изменить. Поэтому, при температуре вблизи критической, переход металл-изолятор вполне может быть спровоцирован изменением концентрации вакансий. Да, линейный закон с изломом может показаться грубым описанием поведения сопротивления вблизи фазового перехода, но для достаточно малой окрестности (по абсолютной концентрации вакансий) точки перехода линейное приближение вполне обосновано и, как минимум, может дать неплохое качественное описание. На практике, для создания мемристоров, работающих в сильно-нелинейном режиме, потребуется довольно точная (однократная для рассматриваемых закрытых мемристоров) подстройка полного количества вакансий и стабилизация температуры.

Конечно, ни о какой сверхпроводимости мемристора (и уж, тем более, его отрицательном сопротивлении) в данном случае речи не идёт. В мемристоре с фазовым переходом концентрация вакансий способна лишь приближаться к предельному значению $c_\mathrm{lim}$, никогда его не достигая. Причина в том, что (при заданном фиксированном токе через мемристор) локальное уменьшение сопротивления уменьшает величину движущего вакансии локального электрического поля. При этом, действие температурных флуктуаций, стремящихся уравнять локальную концентрацию вакансий в соседних точках, не ослабевает. В результате, при сколь угодно большом токе, концентрация вакансий не способна пересечь величину $c_\mathrm{lim}$. Таким образом, в мемристоре с фазовым переходом она лежит уже не в области значений $0 <c <1$, а в интервале $c_\mathrm{lim}<c<1$ при больших положительных $\beta>0$ либо в интервале  $0< c < c_\mathrm{lim}$ при больших (по модулю) отрицательных $\beta<0$. Удельное сопротивление при этом всегда остаётся положительным: $\rho>0$.

\begin{figure}
\begin{center}
\includegraphics[width=1.0\columnwidth]{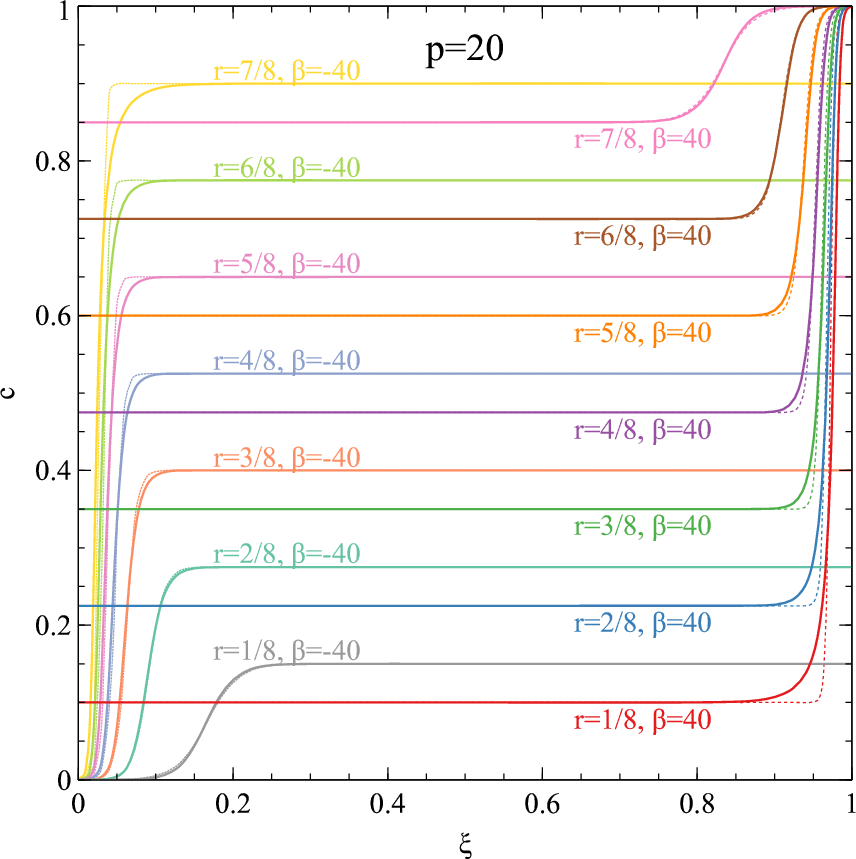}
\vspace{-1cm}
\end{center}
\caption{\label{fig:profilesPhTr} Распределения концентрации вакансий в равновесных состояниях мемристора с фазовым переходом с $p=20$ и $\beta=\pm 40$ при разных заполнениях мемристора; линии --- точное аналитическое выражение~\eqref{eq:parametricC}, точками показано приближение~\eqref{eq:burgersAppr}}
\end{figure}
Примеры равновесных распределений вакансий~\eqref{eq:parametricC} в мемристоре с фазовым переходом приведены для двух значений $\beta$ на Рис.~\ref{fig:profilesPhTr} во всём интервале коэффициентов заполнения мемристора $r$. В принципе, достаточно изобразить лишь один из этих наборов кривых -- при положительных, либо при отрицательных с тем же модулем значениях $\beta$. Второй набор получается из него последовательным применением преобразований~\eqref{eq:betasign}, \eqref{eq:simsignP}, что сводится к повороту всего графика на $180^\circ$ вокруг центральной точки $\xi=1/2$, $c=1/2$ и переобозначению коэффициентов заполнения $r\leftrightarrow (1-r)$.

Можно также заметить, что при сильной нелинейности, концентрация может оказаться ``зажатой'' в  очень узком интервале значений. Для $\beta>0$ это происходит при почти полностью заполненном мемристоре $r\approx 1$, а для $\beta<0$ при почти пустом $r\approx0$. В этой ситуации один из множителей в потоке~\eqref{eq:current} $J \propto c \cdot ( 1 - c)$ перестаёт оказывать существенное влияние на кинетику движения вакансий и его можно заменить средним значением. Тогда и в сильно-нелинейном случае $|\beta|\gg1$ рассматриваемая модель (подобно случаю $\beta=0$) отображается на перемасштабированную модель Бюргерса:
\begin{align}
 \label{eq:burgersAppr}
 c(\xi,\tau) & \approx c_\mathrm{lim} c_\mathrm{B}(- p r \beta, r/c_\mathrm{lim}, \xi,\tau),\ \text{при}\ \beta<0.
\end{align}
Аналогичное выражение для $\beta>0$ получается преобразованием~\eqref{eq:betasign}. Здесь
$c_B (p,r,\xi,\tau)$ --- решение уравнения~\eqref{eq:kinetics} при $\beta=0$, а использованное при построениях графиков $c_\mathrm{B}(p,r,\xi)=\lim_{\tau\rightarrow\infty}c_\mathrm{B}(p,r,\xi,\tau)$ дано уравнением~\eqref{eq:cBurg}. Отметим, что данное отображение справедливо как для стационарных состояний, так и для переходных процессов в мемристоре.

Равновесные состояния, соответствующие~\eqref{eq:burgersAppr}, изображены на Рис.~\ref{fig:profilesPhTr} линиями из точек. Можно заметить, что уже при $r\le1/4$ для $\beta=-40$ (и при $r\ge 3/4$ при $\beta=40$) эти распределения визуально неотличимы от точных~\eqref{eq:parametricC}, показанных сплошными линиями. Важным достоинством отображения~\eqref{eq:burgersAppr} является наличие точных решений не только для равновесных распределений вакансий, но также для зависящих от времени  переходных~\cite{boylo2020} и периодических~\cite{BM2024} процессов. Эти решения автоматически становятся применимы для моделирования таких ``прижатых к фазовому переходу'' сильно-нелинейных мемристоров.

Можно отметить, что с ростом $|\beta|$ большая часть мемристора оказывается заполненной вакансиями почти на уровне предельной концентрации $c_\mathrm{lim}$, то есть находится в состоянии с минимальным значением удельного сопротивления. В рассмотренном случае линейной зависимости удельного сопротивления от концентрации вакансий~\eqref{eq:rho} это никак не отражается на сопротивлении между контактами $\xi=0$ и $\xi=1$. Однако, если добавить третий контакт на некотором промежуточном расстоянии $0<\xi<1$ --- сопротивление между ним и остальными двумя контактами будет сильно зависеть от состояния мемристора. Такое трёх-терминальное устройство под названием мемистор~\cite{Xia2011} уже было предложено и реализовано~\cite{Kaeriyama2005,Wang2013}, в том числе и в контексте нейроморфных вычислений~\cite{BP2017}. Надеемся, что изложенная здесь простая теоретическая модель окажется полезной для развития интуиции, необходимой при оптимизации и проектировании таких устройств.

\section{Выводы}

Рассмотрена модель мемристора из материала с линейной зависимостью удельного сопротивления от концентрации кислородных вакансий (или других заряженных мобильных дефектов). Его кинетика определяется нелинейным уравнением в частных производных, сводящимся в частном случае, когда зависимостью удельного сопротивления от концентрации вакансий можно пренебречь, к уравнению Бюргерса. Получены точные аналитические выражения для распределений вакансий в равновесных состояниях такого мемристора --- пространственных распределений вакансий, созданных долговременным протеканием через него электрического тока при постоянной температуре. В случае слабой (но, всё же, не пренебрежимо малой) зависимости удельного сопротивления от концентрации вакансий, эта зависимость мешает насыщению сопротивления мемристора долговременным пропусканием постоянного электрического тока. Интересно, что рассмотренная простая модель позволяет описывать мемристоры, материал которых претерпевает (в зависимости от концентрации дефектов) фазовый переход металл-изолятор. Наличие перехода сужает допустимый диапазон концентраций мобильных вакансий. Если диапазон становится достаточно узок, кинетику переключения таких мемристоров (и трёхтерминальных мемисторов) можно приближенно отобразить на точно решаемое уравнение Бюргерса с полным учетом линейной зависимости их удельного сопротивления от концентрации вакансий.

\ifarxiv\else\bibliographystyle{jetp}\fi